\documentclass[preprint,epsfig,superscriptaddress]{revtex4}

\usepackage{graphicx}
\usepackage{amssymb}
\usepackage[fleqn]{amsmath}
\usepackage{color}
\usepackage{bm}
\usepackage{physics}
\usepackage{appendix}
\usepackage[export]{adjustbox}
\allowdisplaybreaks

\begin{document}

\title{Engineering high Chern number insulators}

\author{Sungjong Woo}\email{physwoo@gmail.com}
\affiliation{Department of Physics, Pukyong National University, Pusan, Republic of Korea}
\author{Seungbum Woo}\email{seungbum.woo@gmail.com}
\affiliation{Department of Physics and Astronomy, Seoul National University, Seoul, Republic of Korea}
\author{Jung-Wan Ryu}\email{jungwanryu@gmail.com}
\affiliation{Center for Theoretical Physics of Complex Systems, Institute for Basic Science, Daejeon, Republic of Korea}
\author{Hee Chul Park}\email{hc2725@gmail.com}
\affiliation{Department of Physics, Pukyong National University, Pusan, Republic of Korea}

\begin{abstract}
The concept of Chern insulators is one of the most important buliding block of topological physics, enabling the quantum Hall effect without external magnetic fields. 
The construction of Chern insulators has been typically through an guess-and-confirm approach, which can be inefficient and unpredictable. 
In this paper, we introduce a systematic method to directly construct two-dimensional Chern insulators that can provide any nontrivial Chern number. 
Our method is built upon the one-dimensional Rice-Mele model, which is well known for its adjustable polarization properties, providing a reliable framework for manipulation.
By extending this model into two dimensions, we are able to engineer lattice structures that demonstrate predetermined topological quantities effectively. 
This research not only contributes the development of Chern insulators but also paves the way for designing a variety of lattice structures with significant topological implications, potentially impacting quantum computing and materials science. 
With this approach, we are to shed light on the pathways for designing more complex and functional topological phases in synthetic materials.
\end{abstract}

\maketitle

\section{Introduction}

Topological physics, which has been one of the most hot topics in condensed matter physics in the last decade, historically began with quantum Hall effect based on the electronic Landau levels~\cite{K_v_Klitzing}.
Haldane predicted quantum Hall effect without external magnetic field on a graphene, so called a Chern insulator~\cite{F_D_M_Haldane}, which attracted huge attention after topological insulators were found~\cite{C_L_Kane, B_Andrei_Bernevig,M_Z_Hasan}.
Although Haldane's quantum Hall effect on a graphene is much weaker than was predicted, it has its significance in understanding topological materials. 
It is the very basic building block for the theory of band topology.
Differently from the quantum Hall effect under external magnetic field, the Haldane's original model supports Chern number $N_{\rm C}$ only $\pm 1$ or 0.
Chern insulators with Chern number larger than one will be very useful for quantum technology such as quantum computing based on electronics, photonics, and so on.
Graphene-like hexagonal lattice with long range hoppings was theoretically proposed for a high Chern number insulator~\cite{Doru_Sticlet}.
Recently, multilayer quantum anomalous Hall system was also proposed and experimentally realized for high Chern number insulators with various Chern numbers~\cite{Yi-Fan_Zhao, Jun_Ge, Yi-Xiang_Wang, Wenxuan_Zhu}.

Typically, to make a new Chern insulator in a tight-binding model, one has to guess the bulk structure of a system and the complex hopping strengths to find out the Chern number only after system-dependent Berry flux integration.
In this work, we propose a systematic procedure to build Chern insulators with any Chern number.
The procedure is based on the Rice-Mele(RM) model~\cite{M_J_Rice}, a generalized version of the Su-Schrieffer-Heeger(SSH) model~\cite{W_P_Su}.

\section{Theoretical concept}

\begin{figure}[hpbt!]
\centering
\includegraphics[width=0.8\columnwidth]{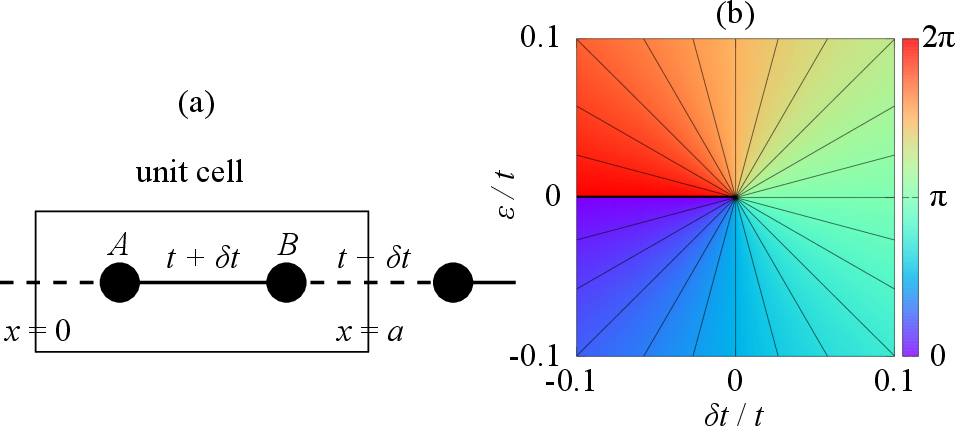}
\caption{(a) The unit cell of Rice-Mele model (b) A colored contour plot of Zak phase of Rice-Mele model as a function of two parameters, on-site energy difference between two subsites, $\varepsilon$, and half of the difference between neigbouring hopping constants, $\delta t$.} 
\label{ZakPhase}
\end{figure}

Figure~\ref{ZakPhase}(a) represents RM model with unit cell size $a$, whose Hamiltonian is
\begin{align}
\begin{aligned}
\label{RM_H}
H=\sum_{R} &\left[{\varepsilon}\left(c_{R,A}^\dagger c_{R,A} - c_{R,B}^\dagger c_{R,B}\right)\right.\\
 &+ \left.(t+\delta t)\left(c_{R,A}^\dagger c_{R,B} + h.c.\right)
  + (t-\delta t)\left(c_{R+a,A}^\dagger c_{R,B} + h.c.\right) \right].
\end{aligned}
\end{align}
Here, $R=na$ represents the cell vectors with $n = \cdots, -2, -1, 0, 1, 2, \cdots$ and $a$ being the cell size. The value of $2\varepsilon$ represents the on-site energy difference between $A$ and $B$ sites, $t+\delta t$ and $t-\delta t$ are inter- and intra-cell hopping constants, $c_{R,A(B)}^\dagger$ is the creation operator on the site $A$($B$) in the unit cell at $R$, and $h.c.$ represents hermitian conjugate.
With Fourier transformation, $c_{R,A(B)}^\dagger = \sqrt\frac{1}{G} \int_0^{G}dk~c_{k,A(B)}^\dagger e^{-ik(R+d_{A(B)})}$, together with positions of two sites within a unit cell, $d_A=\frac{a}{4}$ and $d_B=\frac{3a}{4}$, the Hamiltonian can be rewritten as a Bloch form of $H = \int_0^G dk~\phi_k^\dagger\mathcal{H}(k)\phi_k$ with
\begin{align}
\label{Hk}
\mathcal{H}(k) = 2t\cos \left(\frac{ka}{2}\right) \sigma_1 - 2\delta t\sin \left(\frac{ka}{2}\right) \sigma_2 + {\varepsilon} \sigma_3,
\end{align}
where $G=\frac{2\pi}{a}$ is the reciprocal lattice vector, $\phi^\dagger \equiv (c_{k,A}^\dagger,c_{k,B}^\dagger)$ and $\sigma_i$s are the Pauli matrices (see Appendix for details).
Differently from the SSH model, the Zak phase, $\gamma=i\int_0^G dk~\bra{u(k)}\ket{{\partial_k}u(k)}$~\cite{J_Zak}, can have all the values from 0 to $2\pi$ up to multiple of $2\pi$ phase ambiguity, owing to the energy difference between $A$ and $B$ sites.
Here, $\ket{u(k)}$ is the eigenvector of  $\mathcal{H}(k)$.
We calculate the Zak phase numerically by obtaining $\ket{u(k)}$ and using
\begin{align}
\label{eq_Zak2}
\gamma=-\Im \log \prod_{j=0}^{N-1} \bra{u(k_{j})}\ket{u(k_{j+1})},
\end{align}
instead of its differential form, where, $k_j = \frac{G}{N}j$~\cite{Raffaele_Resta}.
Eq.~\eqref{eq_Zak2} does not need the differentiability of $\ket{u(k)}$ in terms of $k$.
One has to keep in mind that $\ket{u\left(G\right)} = e^{-iG\hat{x}}\ket{u(0)}$ with $\hat{x}$ being the position operator that has two values, $d_A$ or $d_B$.

Figure~\ref{ZakPhase}(b) shows $\gamma$ calculated numerically.
It demonstrates that $\gamma \approx \pi + \arg{\left({2\delta t + i2\varepsilon}\right)}$ for small $\delta t$ and $\varepsilon$ compared to $t$, where $\arg(z)$ means the argument of a complex number $z$.
A pair of parameters can always be found for any given $\gamma$ as
\begin{align}
\label{delt&epsilon}
\delta t \approx -2t'\cos(\gamma),\hspace{.5cm} 
\varepsilon \approx -2t'\sin(\gamma),
\end{align}
with an arbitrary parameter $t' (\ll t)$.
The condition $t'\ll t$, that is required for the approximation of Eq.~\eqref{delt&epsilon}, is in fact not necessary for the topological implication since what only matters is the $2\pi$ phase winding around the sigular point in Fig.~\ref{ZakPhase}(b).
The extra multiples of two are for the simplicity of the formulas in the following part.
Rice-Mele model is a one-dimensional (1D) system that can modulate $\gamma$ with two control parameters, $\delta t$ and $\varepsilon$.
For example, positive $\delta t$ with zero $\varepsilon$ yields $\gamma=\pi$, so that the expectation value of position operator becomes  $\expval{x} = \frac{\gamma}{G} = \frac{a}{2}$; it represents the middle of the strong bond in Fig.~\ref{ZakPhase}(a).
This case corresponds to the trivial topological phase with no edge states.
In many literatures, trivial phase corresponds to $\gamma=0$.
It, however, is only a matter of convention of setting the positions of two subsites $A$ and $B$ while
$\gamma=0$ for trivial topological phase can be achieved from our formalism by setting $d_A=-\frac{a}{4}$ and $d_B=\frac{a}{4}$.
The reason of choice of our convention is to confine two sites within a unit cell.

On the other hand, a Chern insulator is a two-dimensional (2D) system with $\gamma$ along one axis winding the two-torus of the first Brillouin zone with integral number of times.
For a 2D system in a rectangular coordinate system, the matrix representation of a Bloch Hamiltonian can be written as $\mathcal{H}(k_x, k_y)$. 
Let us assume a rectangular lattice system with two orthogonal unit vectors of ${\bf u}_x=(a_x,0)$ and ${\bf u}_y = (0,a_y)$, so that $0\le k_x<G_x(=\frac{2\pi}{a_x})$ and $0\le k_y<G_y(=\frac{2\pi}{a_y}$).
For a fixed $k_x$, one can calculate $\gamma$ along the $y$ axis, $\gamma_y(k_x) = i\int_0^{G_y} dk_y~\bra{u}{{\partial k_y}}\ket{u}$.
If this value varies by $2\pi N_{\rm C}$ as $k_x$ changes by $G_x$, the system is topologically classified to have Chern number of $N_{\rm C}$~\cite{M_Z_Hasan}.

\section{results and discussion}

We propose a procedure to construct a 2D Chern insulator using 1D RM model by building up $2\pi N_{\rm C}$ change of $\gamma$ along an additional dimension of Bloch momentum space;
A 2D system satisfying $\gamma_y(k_x)=N_{\rm C}k_xa_x$ makes $\gamma_y$ change by $2\pi N_{\rm C}$ as $k_x$ changes by $G_x$.
Using Eq.~(\ref{delt&epsilon}), $\delta t$ and $\varepsilon$ for a given $k_x$ can be set to be
\begin{align*}
\delta t = -2t'\cos \left(N_{\rm C}k_xa_x\right),\hspace{.5cm} \varepsilon = -2t'\sin \left(N_{\rm C}k_xa_x\right),
\end{align*}
and hence Eq.~(\ref{Hk}), with $k\rightarrow k_y$, can be extended to 2D Hamiltonian with Chern number $N_{\rm C}$ such that 
$H_{N_{\rm C}} = \int_{\rm BZ}d^2k~\phi_{\bf k}^\dagger\mathcal{H}_{N_{\rm C}}({\bf k})\phi_{\bf k}$,
where $\phi^\dagger \equiv (c_{{\bf k},A}^\dagger,c_{{\bf k},B}^\dagger)$, ${\bf k} = (k_x, k_y)$ and
{
\setlength{\mathindent}{.5cm}
\begin{align}
\label{H_2D_k}
\begin{aligned}
\mathcal{H}_{N_{\rm C}}&({\bf k}) 
=  2t\cos\left(\frac{k_ya_y}{2}\right)\sigma_1
+ 4t'\cos \left(N_{\rm C}k_xa_x\right)\sin\left(\frac{k_ya_y}{ 2}\right)\sigma_2 
-2t'\sin (N_{\rm C}k_xa_x) \sigma_3\\
&=  2t\cos\left(\frac{{\bf k}\cdot{\bf u}_y}{2}\right)\sigma_1
+ 4t'\cos \left(N_{\rm C}{\bf k}\cdot{\bf u}_x\right)\sin\left(\frac{{\bf k}\cdot{\bf u}_y}{ 2}\right)\sigma_2 
-2t'\sin (N_{\rm C}{\bf k}\cdot{\bf u}_x) \sigma_3.
\end{aligned}
\end{align}
}
So is built, this Hamiltonian has $\gamma_y$ varying by the amount of $2\pi{N_{\rm C}}$ as $k_x$ changes by $G_x$; a Chern insulator with a Chern number of $N_{\rm C}$.

In order to find out the corresponding real space representation of this $H_{N_{\rm C}}$, one can use 2D inverse Fourier transformation of the creation operator,
\begin{align}
\label{2D_inv_Fourier}
c_{{\bf k},A(B)}^\dagger = 
{1\over 2\pi}\sum_{\bf R} c_{{\bf R},A(B)}^\dagger 
e^{i{\bf k}\cdot\left({\bf R}+{\bf d}_{A(B)}\right)},
\end{align}
where ${\bf R}=(na_x,ma_y)$ and ${\bf d}_{A(B)} = (0, d_{A(B)})$ with $a_x$ and $a_y$ being the sizes of the unit cells along the $x$ and $y$ axis and $n, m = \cdots, -2, -1, 0, 1, 2, \cdots$.
With $d_A=\frac{a}{4}$ and $d_B=\frac{3a}{4}$, $H_{N_{\rm C}}$ can be transformed back into real space (see Appendix \ref{app1} for details) as
{
\setlength{\mathindent}{0pt}
\begin{align}
\begin{aligned}
    &H_{N_{\rm C}}=t\sum_{{\bf R}}
    \left(c_{{\bf R},A}^\dagger c_{{\bf R},B} + c_{{\bf R}+{\bf u}_y,A}^\dagger c_{{\bf R},B}\right) + h.c.\\
    &+t'\sum_{{\bf R}}
    \left(
    c_{{\bf R}-N_{\rm C}{\bf u}_x,A}^\dagger c_{{\bf R},B}
    - c_{{\bf R}-N_{\rm C}{\bf u}_x+{\bf u}_y,A}^\dagger c_{{\bf R},B}
    + c_{{\bf R}+N_{\rm C}{\bf u}_x,A}^\dagger c_{{\bf R},B}
    - c_{{\bf R}+N_{\rm C}{\bf u}_x+{\bf u}_y,A}^\dagger c_{{\bf R},B}
    \right) + h.c.\\
    &+ it'\sum_{{\bf R}} \left(
    c_{{\bf R}-N_{\rm C}{\bf u}_x,A}^\dagger c_{{\bf R},A}
    - c_{{\bf R}-N_{\rm C}{\bf u}_x,B}^\dagger c_{{\bf R},B}
    \right) + h.c.
\end{aligned}
\end{align}
}

\begin{figure}[hpbt!]
\includegraphics[width=\columnwidth]{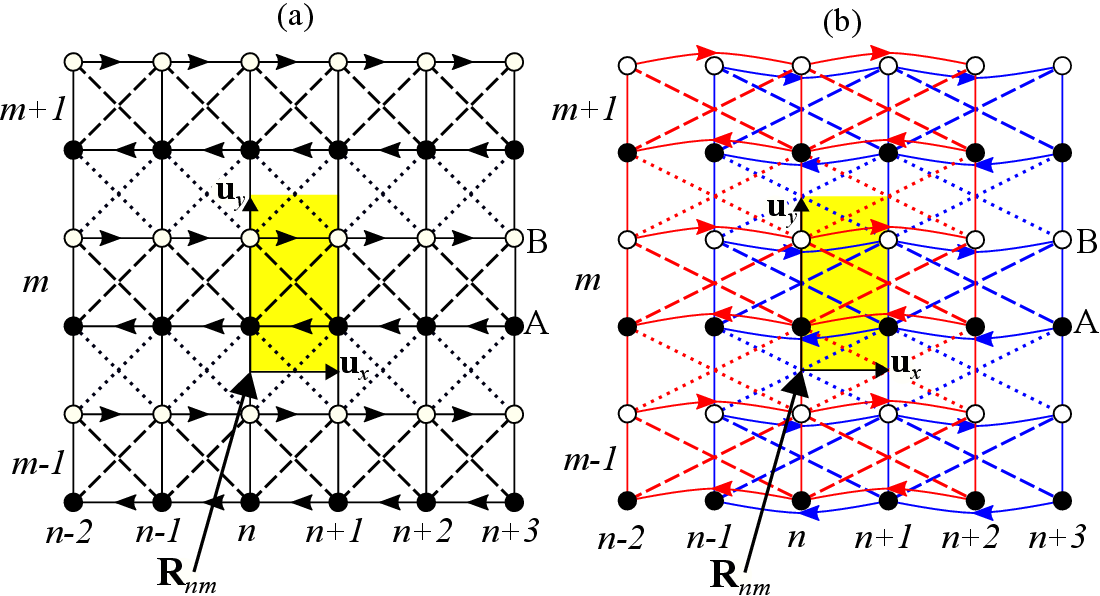}
\caption{The hopping configurations of constructed lattice structures for (a) $N_{\rm C} = 1$ and (b) $N_{\rm C} = 2$. In terms of the hopping configuration, (b) is two copies of (a) that are mutually shifted in space. Find the details in the main text. } 
\label{hopping_C1}
\end{figure}

Figure~\ref{hopping_C1}(a) shows the hopping configuration corresponding to such a system with $N_{\rm C}=1$.
The solid, dashed and dotted lines without arrows represent bondings with real hopping constants of $t$, $t'$ and $-t'$, respectively.
Solid and empty circles repesent $A$ and $B$ sites.
The yellow rectangle is a unit cell with ${\bf u}_x$ and ${\bf u}_y$ as the unit vectors.
The lines with arrows represent hopping constants $it'$ along the direction of the arrow and $-it'$ along the opposite direction.
The convention of hopping direction is that the coefficient of $c_\beta^\dagger c_\alpha$ corresponds to the hopping constant from the $\alpha$ to the $\beta$ site.
It implies staggered magnetic flux perpendicular to the plane which cancels out within each unit cell similar to the Haldane's Chern insulator.
It is noteworthy that the on-site energy difference between two subsites is zero even though it is built from the RM model that has non-zero on-site energy difference.
Instead, the imaginary hopping constants along the $A$ and $B$ sites have opposite signs. 
The inter-site next nearest hopping constants also have two opposite signs within a unit cell.
\begin{figure}[hpbt!]
\includegraphics[width=0.9\columnwidth]{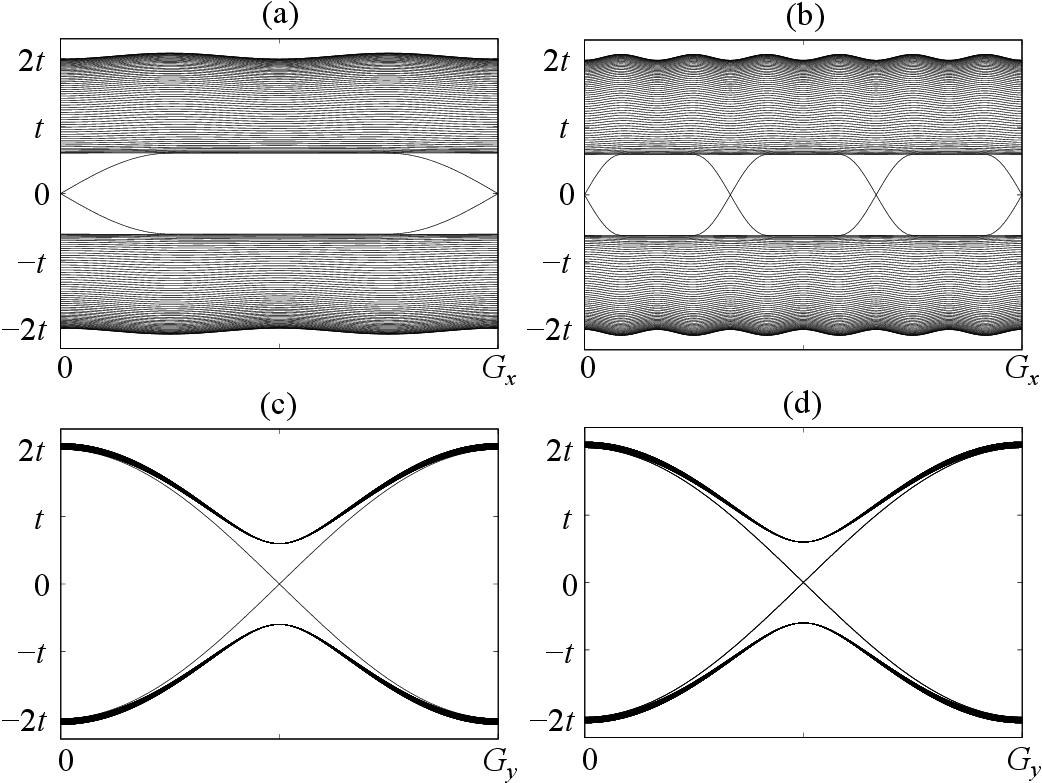}
\caption{Band structure with an open boundary condition with two edges. For the system with $N_C=1$ with edges (a) along the $x$ axis and (c) along the $y$ axis.  For the system with $N_C=3$ with edges (b) along the $x$ axis and (d) along the $y$ axis.  } 
\label{band}
\end{figure}
We have calculated the chiral edge states with open boundary conditions along the $x$ and $y$ axis, respectively. 
Figure~\ref{band}(a) and (c) show the band structures for the case with a pair of edges along the $x$ and $y$ axis respectively for $N_{\rm C}=1$.
The value of $t'$ that decides the gap size is set to be $t'= 0.6t$. The number of unit cells between edges are chosen to be 60.
The number of gapless chiral edge band lines is 2 since there are two edges.

Figure~\ref{hopping_C1}(b) shows the configuration for $N_{\rm C}=2$. 
For a general case of $N_{\rm C}>1$, the $n$-th site is connected to the $(n\pm N_{\rm C})$-th site by hopping along the $x$ axis so that only $(n\pm \eta N_{\rm C})$-th sites  are connected to $n$-th sites with $\eta = 1, 2, 3, \cdots$.
In such a manner, the system is partitioned into $N_{\rm C}$ decoupled subsystems.
In Fig.~\ref{hopping_C1}(b), two subsystems are colored distinctly as red and blue.
Furthermore, each decoupled system is identical to the $N_{\rm C}=1$ system in terms of the hopping graph.
With an open boundary condition, since the structure consists of $N_{\rm C}$ copies of a Chern number one system, it should show $N_{\rm C}$ gapless boundary band lines for each edge. 
This means that its Chern number is $N_{\rm C}$ as supposed.
Figure~\ref{band}(b) and (d) show the band structures for open boundary condition with a pair of edges along the $x$ and $y$ axis respectively, for $N_{\rm C}=3$.
Each of the two band lines corresponding to the edge states on two edges in Fig.~\ref{band}(d) has three-fold degeneracy.

\begin{figure}[hpbt!]
\includegraphics[width=\columnwidth]{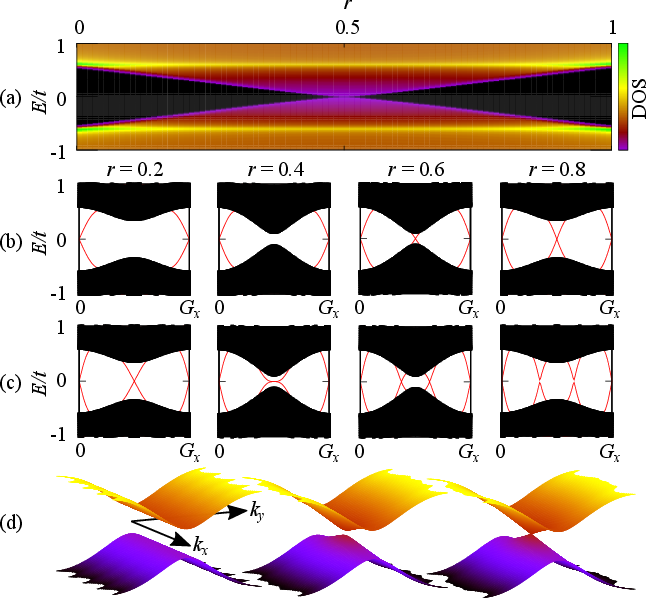}
\caption{(a) Bulk DOS of a mixed system of $N_{\rm C}=1$ and $N_{\rm C}=2$ for various mixing ratio $r$. The horizontal and vertical axes are mixing ratio and energy, respectively.
We confirm that it is identical to that for a mixed system of $N_{\rm C}=2$ and $N_{\rm C}=3$.
(b) Band structures of an open system with varying mixing ratio for a mixed system of $N_{\rm C}=1$ and $N_{\rm C}=2$.
(c) Those for a mixed system of $N_{\rm C}=2$ and $N_{\rm C}=3$.
(d) 2D band structure for the unmixed Chern insulator with $N_{\rm C}=2$ (left), for a mixed one between $N_{\rm C}=1$ and $N_{\rm C}=2$ at the transition point making a Dirac cone (right), for a general mixed one with Chern number two (middle).} 
\label{DOS}
\end{figure}
One might say that this result is trivial since the constructed high-$N_{\rm C}$ Chern insulator is simply an overlap of $N_{\rm C}$ copies of partially translated systems with $N_{\rm C}=1$.
A nontrivial one, however, can further be constructed by mixing Hamiltonians with distinct Chern numbers.
For example, the Hamiltonian that mixes $N_{\rm C}=1$ and $N_{\rm C}=2$ is
\begin{align*}
H_{1,2}(r) = (1-r)H_{N_{\rm C}=1} + rH_{N_{\rm C}=2}
\end{align*}
where $r$ is the mixing ratio.
If $r=0$ the Hamiltonian becomes one with $N_{\rm C}=1$ while if $r=1$ it becomes one with $N_{\rm C}=2$.
Therefore, as $r$ varies from 0 to 1, the Chern number of $H_{1,2}(r)$ must undergo at least one topological phase transition.
At the topological transition, the bulk band gap must be closed.
Figure~\ref{DOS}(a) shows the bulk density of state (DOS) as a function of $r$ demonstrating gap closing at $r=0.5$; $N_{\rm C}=1$ for $r<0.5$ and $N_{\rm C}=2$ for $r>0.5$.
Figure~\ref{DOS}(b) shows how the number of gapless band lines increases.
Further numerical calculations confirm that the change of DOS as a funtion of the mixing ratio from $N_{\rm C}=2$ to $3$ with $H_{2,3}(r)$ is identical to the case from $N_{\rm C}=1$ to $2$.
Figure~\ref{DOS}(c) shows the discrete change of number of gap closing band lines from 4 to 6 as $r$ changes; $N_{\rm C}$ makes a transition from 2 to 3 at $r=0.5$.
Figure~\ref{DOS}(d) shows the 2D band structure for the unmixed Chern insulator with $N_{\rm C}=2$ (left), for a mixed one between $N_{\rm C}=1$ and $2$ at the transition point which closes gap making a Dirac cone (right), and for a general mixed one with Chern number two that shows quadratic dispersion at the zone boundary (middle). 
This demonstrates that by mixing different unmixed Hamiltonians, $H_{N_{\rm C}}$, with different $N_{\rm C}$ values in a various way, nontrivial high Chern number insulators can be constructed on demand.

\begin{figure}[hpbt!]
\includegraphics[width=0.8\columnwidth]{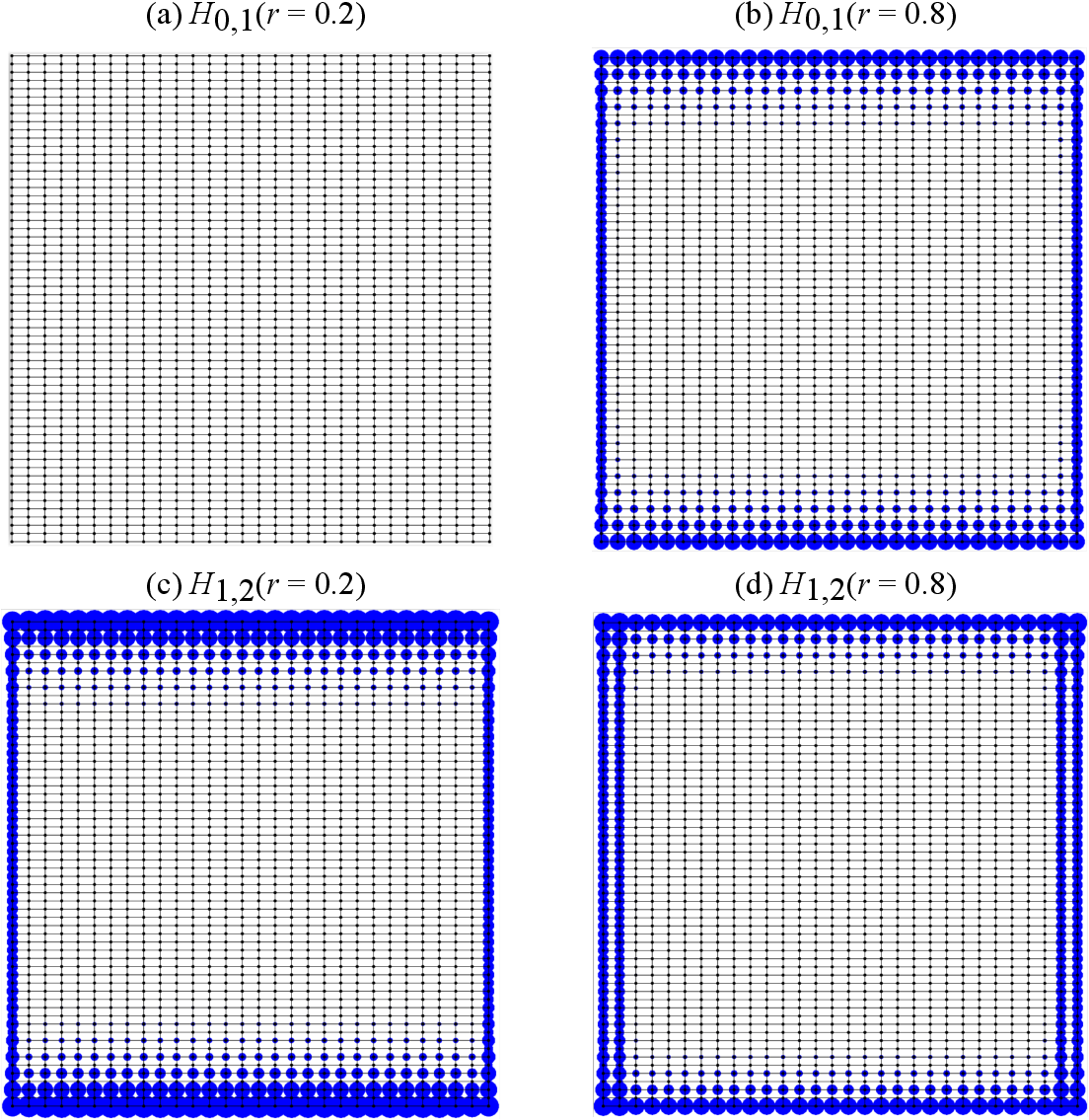}
\caption{$H_{\alpha,\beta}(r) = (1-r)H_{N_{\rm C}=\alpha}+rH_{N_{\rm C}=\beta}$ where $H_{N_{\rm C}}$ represents the Hamiltonian of a pure system with the Chern number of $N_{\rm C}$. The ldos at $E=0$ within the bulk energy gap with energy width of $0.02t$ for isolated systems with $30\times30$ unit cells. (a) Since Chern number is zero there is no gap states. The ldos within the gap is zero. (b) and (c) demonstrate the chrial edge states for Chern number one. (d) corresponds to Chern number two.} 
\label{ldos}
\end{figure}
We further calculate local density of states (ldos) at $E=0$ with energy width of $0.02t$ for isolated systems of size $30\times30$ unit cells with an open boundary condition by varying the value of $r$ for $H_{0,1}(r)$ and $H_{1,2}(r)$ to demonstrate the emergence of chiral edge states.
Figure~\ref{ldos} shows zero-energy ldos showing gap state density for (a) $H_{0,1}(r=0.2)$, (b) $H_{0,1}(r=0.8)$, (c) $H_{1,2}(r=0.2)$, and (d) $H_{1,2}(r=0.8)$. 
It demonstrates that the chiral edge states emerge once $r$ becomes larger than 0.5 in the case with $H_{0,1}(r)$~[Figs.~\ref{ldos}(a, b)].
Figures~\ref{ldos}(b, c) have Chern number of one.
Figure~\ref{ldos}(d), that corresponds to Chern number two, shows that the edge state density covers two rows on each edge along the $y$ axis, differently from (b) and (c) where it covers only one row.
Although not being a general manefestation of Chern number, it can be used for a signature of topological phase transition for this specific system if tested experimentally in the future.

\begin{figure}[hpbt!]
\includegraphics[width=1\columnwidth]{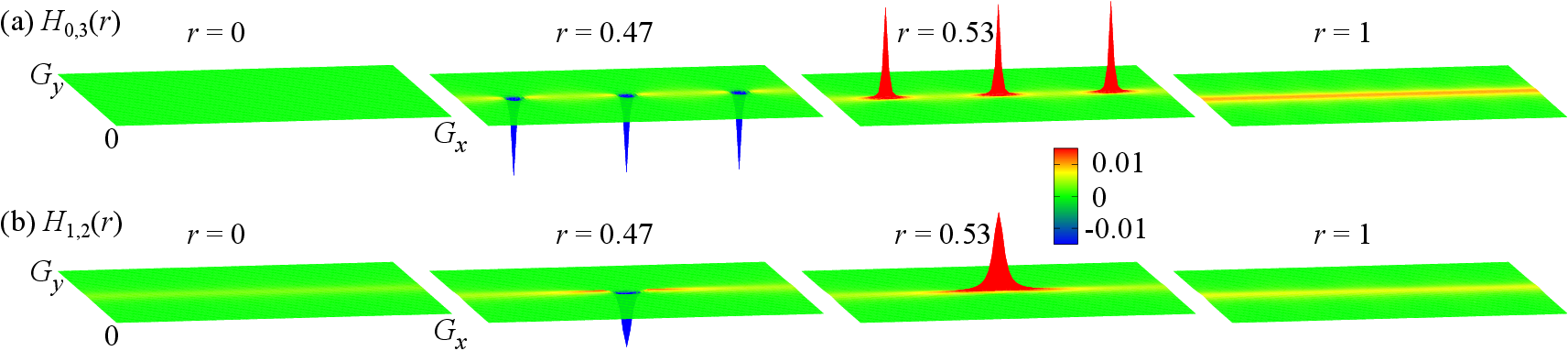}
\caption{Berry flux in the first Brillouin zone. (a) Mixed hamiltonian of $N_{\rm C}=0$ and $N_{\rm C}=3$ with mixing ratio $r=0, 0.47, 0.53, 1$. The topological phase transition occurs at $r=0.5$ from Chern number 0 to 3. (b) Mixed hamiltonian of $N_{\rm C}=1$ and $N_{\rm C}=2$ with the same mixing ratios. The topological phase transition also occurs at $r=0.5$ from Chern number 1 to 2.} 
\label{BerryFlux}
\end{figure}
In order to further confirm the Chern number, we calculate Berry flux within the first Brillouin zone across the phase transition.
Figure~\ref{BerryFlux}(a) and (b) shows the Berry flux for systems with $H_{0,3}(r)$ and $H_{1,2}(r)$ for different values of $r$.
Both cases undergo phase transition at $r=0.5$.
For $H_{0,3}(r)$, Chern number changes directly from $0$ to $3$ at $r=0.5$. 
The second and third plots are Berry flux right before and after phase transtion with Chern numbers, $0$ and $3$.
The integration of Berry flux over the Brillouin zone divided by $2\pi$ is the Chern number.
For the case before the phase transition with $r=0.47$, integration of the dispersed positive Berry flux (yellow) exactly cancels out with that of the three localized spots of negative Berry flux (blue) to make the Chern number zero.
On the other hand, for the case after the phase transition with $r=0.53$, the localized spots of Berry flux becomes singular when $r=0.5$, changing their sign of flux to plus making the Chern number three.
For $H_{1,2}(r)$, the story is similar with phase transition from Chern number one to two through one singular spot of Berry flux.
We confirm that singular points of Berry flux correspond to gap closing points.

Lastly, we would like to mention briefly about experimental realization and future plans. 
As for the experimental realization, due to its relatively complicated nature of hopping structures it might be challenging in solid state systems.
However, complicated hopping structures have been widely achieved recently in the fields such as optics and electric circuits. 
The next nearest neighbor and long-range hopping can be implemented in photonic systems such as an array of coupled optical-ring resonators and coupled-resonator optical waveguides, which consist of site resonators and link resonators \cite{M_Hafezi,Sunil_Mittal,Daniel_Leykam,JungYun_Han}. 
In a synthetic dimension, long-range hopping can also be implemented by utilizing frequency modes in two coupled ring resonators \cite{Kai_Wang}. 
Asymmetric long-range hopping between resonant elements in a photonic arrangement have been achieved in active optical platforms \cite{Yuzhou_G_N_Liu} and complex hopping between neighboring lattice sites have been realized in electronic circuit layouts \cite{Jia_Ningyuan,Xiang_Ni,Lingling_Song}.
Our future work will include investigation of dynamical responses of topological edges states to varying Chern number by tuning the hopping parameters of these systems.
It is also important to note that the experimental realization of high Chern number insulators are mostly multilayered systems~\cite{Yi-Fan_Zhao, Jun_Ge, Yi-Xiang_Wang, Wenxuan_Zhu}.
They consist of layers each of which has Chern number one while multilayers make copies of each layer introducing high Chern number.
Such mechanism is analogous to our structure; our systems consist of laterally shifted copies of Chern number one systems.


\section{Conclusion}

In conclusion, we propose a systematic method to construct a two-dimensional lattice structure with proper hopping parameters that can give an arbitrary Chern number, based on extended 1D Rice-Mele model. 
With constructed structures, we successfully demonstrate topological phase transitions between different Chern numbers by modulating hopping parameters.
We also demonstrate the emergence of topological chiral edge states across the phase transition.
Generalization of our method to different lattice structures such as hexagonal lattice as well as extension to higher dimensions for higher-order topological insulators shall be of interest for future projects.

\section{acknowledgements}

This work was supported by the National Research Foundation of Korea(NRF) grant funded by the Korea government(MSIT) (No. RS-2023-00278511), by the Pukyong National University Research Fund in (No.202303660001), by Korea Institute for Advanced Study (KIAS), and by the Institute for Basic Science in Korea (No. IBS-R024-D1).

\section{references}

\bibliography{HighChernNumber}

\appendix
\section{Fourier Transformation}
\label{app0}
As in Eq.~\eqref{RM_H}, the Hamiltonian of a RM model is

\begin{align}
\begin{aligned}
\label{eq:RM_H_App}
H=\sum_{R} &\left[{\varepsilon}\left(c_{R,A}^\dagger c_{R,A} - c_{R,B}^\dagger c_{R,B}\right)\right.\\
 &+ \left.(t+\delta t)\left(c_{R,A}^\dagger c_{R,B} + h.c.\right)
 + (t-\delta t)\left(c_{R+a,A}^\dagger c_{R,B} + h.c.\right) \right].
\end{aligned}
\end{align}
With Fourier transformation, $c_{R,A(B)}^\dagger = \sqrt\frac{1}{G} \int_0^{G}dk~c_{k,A(B)}^\dagger e^{-ik(R+d_{A(B)})}$, each term can be transformed into $k$-space.
The term of $\sum c_{R,A}^\dagger c_{R,A}$ is transformed as
\begin{align}
\label{eq:RM_H_App1}
  \begin{aligned}
    \sum_R &c_{R,A}^\dagger c_{R,A} \\
    &= \frac{1}{G}\sum_R 
    \int^G_0 dk~c_{k,A}^\dagger e^{-ik(R+d_{A})}~
    \int^G_0 dk'~c_{k',A} e^{ik'(R+d_{A})}\\
    &= \iint^G_0 dk~dk'~
    c_{k,A}^\dagger c_{k',A}~
    \frac{1}{G} \sum_R e^{-i(k-k')R} \\
    &= \iint^G_0 dk~dk'~
    c_{k,A}^\dagger c_{k',A}~
    \delta(k-k') \\
    &= \iint^G_0 dk~
    c_{k,A}^\dagger c_{k,A}.
  \end{aligned}
\end{align}
The term, $\sum c_{R,B}^\dagger c_{R,B}$ is transformed similarly as
\begin{align}
\label{eq:RM_H_App2}
  \begin{aligned}
    \sum_R &c_{R,B}^\dagger c_{R,B} = \iint^G_0 dk~
    c_{k,B}^\dagger c_{k,B}.
  \end{aligned}
\end{align}
The term, $\sum c_{R,A}^\dagger c_{R,B}$ is transformed as
\begin{align}
\nonumber
  \begin{aligned}
    \sum_R &c_{R,A}^\dagger c_{R,B} \\
    &= \frac{1}{G}\sum_R 
    \int^G_0 dk~c_{k,A}^\dagger e^{-ik(R+d_{A})}
    \int^G_0 dk'~c_{k',B} e^{ik'(R+d_{B})}\\
    &= \iint^G_0 dk~dk'~
    c_{k,A}^\dagger c_{k',B}~
    e^{-ikd_A + ik'd_B}~
    \frac{1}{G} \sum_R e^{-i(k-k')R} \\
    &= \iint^G_0 dk~dk'~
    c_{k,A}^\dagger c_{k',B}~
    e^{-ikd_A + ik'd_B}~
    \delta(k-k') \\
    &= \int^G_0 dk~
    c_{k,A}^\dagger c_{k,B}~
    e^{-ik(d_A-d_B)}.
  \end{aligned}
\end{align}
With $d_A=\frac{a}{4}$ and $d_B=\frac{3a}{4}$, it becomes
\begin{align}
\label{eq:RM_H_App3}
  \sum_R &c_{R,A}^\dagger c_{R,B} 
  = \int^G_0 dk~c_{k,A}^\dagger c_{k,B}~e^{i\frac{ka}{2}}.
\end{align}
The term, $\sum c_{R+a,A}^\dagger c_{R,B}$ is transformed as
\begin{align}
\nonumber
  \begin{aligned}
    \sum_R &c_{R+a,A}^\dagger c_{R,B} \\
    &= \frac{1}{G}\sum_R 
    \int^G_0 dk~c_{k,A}^\dagger e^{-ik(R+a+d_{A})}
    \int^G_0 dk'~c_{k',B} e^{ik'(R+d_{B})}\\
    &= \iint^G_0 dk~dk'~
    c_{k,A}^\dagger c_{k',B}~
    e^{-ik(a+d_A) + ik'd_B}~
    \frac{1}{G} \sum_R e^{-i(k-k')R} \\
    &= \iint^G_0 dk~dk'~
    c_{k,A}^\dagger c_{k',B}~
    e^{-ik(a+d_A) + ik'd_B}~
    \delta(k-k') \\
    &= \int^G_0 dk~
    c_{k,A}^\dagger c_{k,B}
    e^{-ik(a+d_A-d_B)}.
  \end{aligned}
\end{align}
With $d_A=\frac{a}{4}$ and $d_B=\frac{3a}{4}$, it becomes
\begin{align}
\label{eq:RM_H_App4}
  \sum_R &c_{R+a,A}^\dagger c_{R,B} 
  = \int^G_0 dk~c_{k,A}^\dagger c_{k,B}~e^{-i\frac{ka}{2}}.
\end{align}
Summing up terms from Eqs.~(\ref{eq:RM_H_App1}, \ref{eq:RM_H_App2}, \ref{eq:RM_H_App3}, \ref{eq:RM_H_App4}), Eq.~\eqref{eq:RM_H_App} can be represented in $k$-space as $H = \int^G_0 dk~\mathcal{H}(k)$ with
\begin{align}
  \begin{aligned}
    \mathcal{H}=&\varepsilon
    \left( c_{k,A}^\dagger c_{k,A} - c_{k,B}^\dagger c_{k,B}\right)\\
    &+(t+\delta t)
    \left( c_{k,A}^\dagger c_{k,B}~e^{i\frac{ka}{2}} + h.c.\right) 
    +(t-\delta t)
    \left( c_{k,A}^\dagger c_{k,B}~e^{-i\frac{ka}{2}} + h.c.\right)\\
    =& \varepsilon
    \left( c_{k,A}^\dagger c_{k,A} - c_{k,B}^\dagger c_{k,B}\right) 
    +2t \left(c_{k,A}^\dagger c_{k,B} + c_{k,B}^\dagger c_{k,A}\right) \cos{\left(\frac{ka}{2}\right)}+\\
    &+2i\delta t \left(c_{k,A}^\dagger c_{k,B} - c_{k,B}^\dagger c_{k,A}\right) \sin{\left(\frac{ka}{2}\right)} \\
    =& 
    \begin{pmatrix}
      c^\dagger_{k,A} & c^\dagger_{k,B}   
    \end{pmatrix}
    \begin{pmatrix}
      \varepsilon & 2t\cos\left(\frac{ka}{2}\right)+2i\delta t\sin\left(\frac{ka}{2}\right)\\
      2t\cos\left(\frac{ka}{2}\right)-2i\delta t\sin\left(\frac{ka}{2}\right) & -\varepsilon
    \end{pmatrix}
    \begin{pmatrix}
      c_{k,A} \\ c_{k,B}   
    \end{pmatrix} \\
    =& 
    \phi^\dagger_{k}
    \left(
    2t\cos\left(\frac{ka}{2}\right)
    \sigma_1
    - 2\delta t\sin\left(\frac{ka}{2}\right)
    \sigma_2
    + \varepsilon
    \sigma_3
    \right)
    \phi_{k},
  \end{aligned}
\end{align}
where $\phi^\dagger \equiv (c_{k,A}^\dagger,c_{k,B}^\dagger)$ and $\sigma_i$s are Pauli matrices.

\section{Inverse Fourier Transformation}
\label{app1}
From Eq.~\eqref{H_2D_k}, the hamiltonian with Chern number $N_{\rm C}$ in the $\bf k$-space representation consists of three terms as
\begin{subequations}
  \label{eq:HNC}
  \begin{align}
    H_{N_{\rm C}} =& \int_{\rm BZ}d^2k~\phi_{\bf k}^\dagger\mathcal{H}_{N_{\rm C}}({\bf k})\phi_{\bf k}\nonumber\\
    \label{eq:HNC_1}
    =&+2t\int_{\rm BZ} d^2k  \cos\left({{\bf k}\cdot{\bf a}\over 2}\right)\phi_{\bf k}^\dagger\sigma_1\phi_{\bf k}\\
    \label{eq:HNC_2}
    &+ 4t'\int_{\rm BZ} d^2k \cos \left(N_{\rm C}{\bf k}\cdot{\bf b}\right)\sin\left({{\bf k}\cdot{\bf a}\over 2}\right)\phi_{\bf k}^\dagger\sigma_2\phi_{\bf k}\\
    \label{eq:HNC_3}
    &-2t'\int_{\rm BZ} d^2k \sin (N_{\rm C}{\bf k}\cdot{\bf b}) \phi_{\bf k}^\dagger\sigma_3\phi_{\bf k}.
  \end{align}  
\end{subequations}
Each term can be transformed into the representation of real space using Eq.~\eqref{2D_inv_Fourier}.
The first term, Eq.~\eqref{eq:HNC_1}, transforms as
\begin{align}
  \label{eq:HNC1}
  \begin{aligned}
    2t\int_{\rm BZ} &d^2k~\cos\left({{\bf k}\cdot{\bf a}\over 2}\right)\phi_{\bf k}^\dagger\sigma_1\phi_{\bf k}\\
    &=2t\int_{\rm BZ} d^2k~
    c_{{\bf k},A}^\dagger c_{{\bf k},B}
    + h.c.\\
    &=2t\int_{\rm BZ} d^2k~\cos\left({{\bf k}\cdot{\bf a}\over 2}\right)
    {1\over (2\pi)^2}\sum_{{\bf R}, {\bf R'}}
    c_{{\bf R},A}^\dagger e^{i{\bf k}\cdot({\bf R}+{\bf d}_A)}
    c_{{\bf R}',B} e^{-i{\bf k}\cdot({\bf R}'+{\bf d}_B)} + h.c.\\
    &=2t\sum_{{\bf R}, {\bf R'}}
    c_{{\bf R},A}^\dagger c_{{\bf R}',B}
    {1\over (2\pi)^2}\int_{\rm BZ} d^2k~\cos\left({{\bf k}\cdot{\bf a}\over 2}\right)
    e^{i{\bf k}\cdot({\bf R}-{\bf R}'-\frac{\bf b}{2})} + h.c.\\
    &=t\sum_{{\bf R}, {\bf R'}}
    c_{{\bf R},A}^\dagger c_{{\bf R}',B}
    {1\over (2\pi)^2}\int_{\rm BZ} d^2k~\left(e^{i\frac{{\bf k}\cdot{\bf a}}{2}}+e^{-i\frac{{\bf k}\cdot{\bf a}}{2}}\right)
    e^{i{\bf k}\cdot({\bf R}-{\bf R}'-{\frac{\bf a}{2}})} + h.c.\\
    &=t\sum_{{\bf R}, {\bf R'}}
    c_{{\bf R},A}^\dagger c_{{\bf R}',B}
    {1\over (2\pi)^2}\int_{\rm BZ} d^2k~\left(e^{i{\bf k}\cdot\left({\bf R}-{\bf R}'\right)}+e^{i{\bf k}\cdot({\bf R}-{\bf R}'-{\bf a})}\right)
     + h.c.\\
    &=t\sum_{{\bf R}, {\bf R'}}
    c_{{\bf R},A}^\dagger c_{{\bf R}',B}
    \left(\delta_{{\bf R}, {\bf R}'} + \delta_{{\bf R}, {\bf R}'+{\bf a}} \right)
     + h.c.\\
    &=t\sum_{{\bf R}}
    \left(c_{{\bf R},A}^\dagger c_{{\bf R},B} + c_{{\bf R}+{\bf a},A}^\dagger c_{{\bf R},B}\right)
     + h.c.
 \end{aligned}
\end{align}
Here, ${\bf d}_B - {\bf d}_A = {\bf{a}\over 2}$ is used.
The sencond term, Eq.~\eqref{eq:HNC_2}, transforms as
{
\setlength{\mathindent}{0pt}
\begin{align}
  \label{eq:HNC2}
  \begin{aligned}
    4&t'\int_{\rm BZ} d^2k~\cos\left(N_{\rm C}{\bf k}\cdot{\bf b}\right)\sin\left({{\bf k}\cdot{\bf a}\over 2}\right)\phi_{\bf k}^\dagger\sigma_2\phi_{\bf k}\\
    &=4t'\int_{\rm BZ} d^2k~\cos\left(N_{\rm C}{\bf k}\cdot{\bf b}\right)\sin\left({{\bf k}\cdot{\bf a}\over 2}\right)
    \left(-ic_{{\bf k},A}^\dagger c_{{\bf k},B}\right)
    + h.c.\\
    &=4t'\int_{\rm BZ} d^2k~\cos\left(N_{\rm C}{\bf k}\cdot{\bf b}\right)\sin\left({{\bf k}\cdot{\bf a}\over 2}\right)
    {-i\over (2\pi)^2}\sum_{{\bf R}, {\bf R'}}
    c_{{\bf R},A}^\dagger e^{i{\bf k}\cdot({\bf R}+{\bf d}_A)}
    c_{{\bf R}',B} e^{-i{\bf k}\cdot({\bf R}'+{\bf d}_B)} + h.c.\\
    &=4t'\sum_{{\bf R}, {\bf R'}}
    c_{{\bf R},A}^\dagger c_{{\bf R}',B}
    {-i\over (2\pi)^2}\int_{\rm BZ} d^2k~\cos\left(N_{\rm C}{\bf k}\cdot{\bf b}\right)\sin\left({{\bf k}\cdot{\bf a}\over 2}\right)
    e^{i{\bf k}\cdot({\bf R}-{\bf R}'-\frac{\bf a}{2})} + h.c.\\
    &=-t'\sum_{{\bf R}, {\bf R'}}
    c_{{\bf R},A}^\dagger c_{{\bf R}',B}
    {1\over (2\pi)^2}\int_{\rm BZ} d^2k~
    \left(e^{iN_{\rm C}{{\bf k}\cdot{\bf b}}}+e^{-iN_{\rm C}{{\bf k}\cdot{\bf b}}}\right)
    \left(e^{i\frac{{\bf k}\cdot{\bf a}}{2}}-e^{-i\frac{{\bf k}\cdot{\bf a}}{2}}\right)
    e^{i{\bf k}\cdot({\bf R}-{\bf R}'-{\frac{\bf a}{2}})} + h.c.\\
    &=-t'\sum_{{\bf R}, {\bf R'}}
    c_{{\bf R},A}^\dagger c_{{\bf R}',B}\\
    &\hspace{.5cm}\times
    {1\over (2\pi)^2}\int_{\rm BZ} d^2k~
    \left(
    e^{i{\bf k}\cdot\left(N_{\rm C}{\bf b}\right)} -
    e^{i{\bf k}\cdot\left(N_{\rm C}{\bf b}-{\bf a}\right)} +
    e^{i{\bf k}\cdot\left(-N_{\rm C}{\bf b}\right)} -
    e^{i{\bf k}\cdot\left(-N_{\rm C}{\bf b}-{\bf a}\right)}
    \right)
    e^{i{\bf k}\cdot({\bf R}-{\bf R}')} + h.c.\\
    &=-t'\sum_{{\bf R}, {\bf R'}}
    c_{{\bf R},A}^\dagger c_{{\bf R}',B}
    \left(
    \delta_{{\bf R}, {\bf R}'-N_{\rm C}{\bf b}} -
    \delta_{{\bf R}, {\bf R}'-N_{\rm C}{\bf b}+{\bf a}} +
    \delta_{{\bf R}, {\bf R}'+N_{\rm C}{\bf b}} -
    \delta_{{\bf R}, {\bf R}'+N_{\rm C}{\bf b}+{\bf a}}
    \right) + h.c.\\
    &=-t'\sum_{{\bf R}}
    \left(
    c_{{\bf R}-N_{\rm C}{\bf b},A}^\dagger c_{{\bf R},B}
    - c_{{\bf R}-N_{\rm C}{\bf b}+{\bf a},A}^\dagger c_{{\bf R},B}
    + c_{{\bf R}+N_{\rm C}{\bf b},A}^\dagger c_{{\bf R},B}
    - c_{{\bf R}+N_{\rm C}{\bf b}+{\bf a},A}^\dagger c_{{\bf R},B}
    \right)
    + h.c.
  \end{aligned}
\end{align}
}
The third term, Eq.~\eqref{eq:HNC_3} transforms as
\begin{align}
  \label{eq:HNC3}
  \begin{aligned}
    -2t'&\int_{\rm BZ} d^2k~\sin\left(N_{\rm C}{\bf k}\cdot{\bf b}\right)\phi_{\bf k}^\dagger\sigma_3\phi_{\bf k}\\
    &=-2t'\int_{\rm BZ} d^2k~\sin\left(N_{\rm C}{\bf k}\cdot{\bf b}\right)
    \left(c_{{\bf k},A}^\dagger c_{{\bf k},A}-c_{{\bf k},B}^\dagger c_{{\bf k},B}\right)\\
    &=-2t'\int_{\rm BZ} d^2k~\sin\left(N_{\rm C}{\bf k}\cdot{\bf b}\right)
    {1\over (2\pi)^2}\sum_{{\bf R}, {\bf R'}}
    \left(c_{{\bf R},A}^\dagger c_{{\bf R}',A} e^{i{\bf k}\cdot({\bf R}-{\bf R}')}
    - c_{{\bf R},B}^\dagger c_{{\bf R}',B} e^{i{\bf k}\cdot({\bf R}-{\bf R}')}\right)\\
    &=-2t'\sum_{{\bf R}, {\bf R'}}
    \left(c_{{\bf R},A}^\dagger c_{{\bf R}',A}
    - c_{{\bf R},B}^\dagger c_{{\bf R}',B}\right)
    {1\over (2\pi)^2}\int_{\rm BZ} d^2k~\sin\left(N_{\rm C}{\bf k}\cdot{\bf b}\right)
    e^{i{\bf k}\cdot({\bf R}-{\bf R}')}\\
    &=it'\sum_{{\bf R}, {\bf R'}}
    \left(c_{{\bf R},A}^\dagger c_{{\bf R}',A}
    - c_{{\bf R},B}^\dagger c_{{\bf R}',B}\right)
    {1\over (2\pi)^2}\int_{\rm BZ} d^2k~
    \left(
    e^{i{\bf k}\cdot N_{\rm C}{\bf b}}-
    e^{-i{\bf k}\cdot N_{\rm C}{\bf b}}
    \right)
    e^{i{\bf k}\cdot({\bf R}-{\bf R}')}\\
    &=it'\sum_{{\bf R}, {\bf R'}}
    \left(c_{{\bf R},A}^\dagger c_{{\bf R}',A}
    - c_{{\bf R},B}^\dagger c_{{\bf R}',B}\right)
    \left(
    \delta_{{\bf R}, {\bf R}' - N_{\rm C}{\bf b}} - 
    \delta_{{\bf R}, {\bf R}' + N_{\rm C}{\bf b}}
    \right)\\
    &=it'\sum_{{\bf R}} \left(
    c_{{\bf R}-N_{\rm C}{\bf b},A}^\dagger c_{{\bf R},A}
    - c_{{\bf R}+N_{\rm C}{\bf b},A}^\dagger c_{{\bf R},A}
    - c_{{\bf R}-N_{\rm C}{\bf b},B}^\dagger c_{{\bf R},B}
    + c_{{\bf R}+N_{\rm C}{\bf b},B}^\dagger c_{{\bf R},B}
    \right) \\
    &=it'\sum_{{\bf R}} \left(
    c_{{\bf R}-N_{\rm C}{\bf b},A}^\dagger c_{{\bf R},A}
    - c_{{\bf R}-N_{\rm C}{\bf b},B}^\dagger c_{{\bf R},B}
    \right) + h.c.
  \end{aligned}
\end{align}
Summing up terms from Eqs.~(\ref{eq:HNC1}, \ref{eq:HNC2}, \ref{eq:HNC3}), the Hamiltonian can be written in the real space as
{
\setlength{\mathindent}{0pt}
\begin{align}
\begin{aligned}
    H_{N_{\rm C}}&=t\sum_{{\bf R}}
    \left(c_{{\bf R},A}^\dagger c_{{\bf R},B} + c_{{\bf R}+{\bf a},A}^\dagger c_{{\bf R},B}\right) + h.c.\\
    &+\frac{t'}{2}\sum_{{\bf R}}
    \left(
    c_{{\bf R}-N_{\rm C}{\bf b},A}^\dagger c_{{\bf R},B}
    - c_{{\bf R}-N_{\rm C}{\bf b}+{\bf a},A}^\dagger c_{{\bf R},B}
    + c_{{\bf R}+N_{\rm C}{\bf b},A}^\dagger c_{{\bf R},B}
    - c_{{\bf R}+N_{\rm C}{\bf b}+{\bf a},A}^\dagger c_{{\bf R},B}
    \right) + h.c.\\
    &+ it'\sum_{{\bf R}} \left(
    c_{{\bf R}-N_{\rm C}{\bf b},A}^\dagger c_{{\bf R},A}
    - c_{{\bf R}-N_{\rm C}{\bf b},B}^\dagger c_{{\bf R},B}
    \right) + h.c.
\end{aligned}
\end{align}
}

\end{document}